\documentstyle[12pt]{article}
\newlength{\abstractwidth}
\def \be {\begin{equation}}
\def \ee {\end{equation}}
\def \bea {\begin{eqnarray}}
\def \eea {\end{eqnarray}}
\def \nn {\nonumber}
\def \ba {\begin{array}}
\def \ea {\end{array}}

\def \a {\alpha}
\def \b {\beta}
\def \g {\gamma}

\def \d {\delta}

\def \k {\kappa}
\def \lam {\lambda}

\def \cNt {{\cal N}_{\theta}}
\def \cCe {{\cal C}_e}
\def \cCo {{\cal C}_o}
\def \cDe {{\cal D}_e}
\def \cDo {{\cal D}_o}
\def \cMp  {{\cal M}^{'}}
\def \cM {{\cal M}}
\def \cR  {{\cal R}}
\def \cV  {{\cal V}}
\def \cK  {{\cal K}}
\begin{document}
\begin{titlepage}
\bigskip
\bigskip\bigskip\bigskip\bigskip
\bigskip \bigskip
\centerline{\Large \bf {Star Spectroscopy
 in  the Constant B field Background}}
\bigskip\bigskip
\bigskip\bigskip
\centerline{Bin Chen$^{*}$\footnote{chenb@ictp.trieste.it}  and
Feng-Li Lin $^{\dagger}$\footnote{linfl@mail.tku.edu.tw}}
\bigskip
\centerline{\it ${}^*$High Energy Section} \centerline{\it the
Abdus Salam ICTP} \centerline{\it Strada Costiera, 11}
\centerline{\it 34014 Trieste, Italy}
\bigskip
\centerline{\it ${}^\dagger$Department of Physics }
\centerline{\it Tamkang University} \centerline{\it Tamsui, Taipei
25137, Taiwan }
\bigskip\bigskip

%ABSTRACT

\begin{abstract}
\medskip
In this paper we calculate the spectrum of Neumann matrix with
zero modes in the presence of the constant B field in Witten's
cubic string field theory. We find both the continuous spectrum
inside $[{-1\over3}, 0)$ and the constraint on the existence of
the discrete spectrum. For generic $\theta$, $-1/3$ is not in the
discrete spectrum but in the continuous spectrum.  For each
eigenvalue in the continuous spectrum there are four
twist-definite degenerate eigenvectors except for $-1/3$ at which
the degeneracy is two. However, for each twist-definite
eigenvector the twist parity is opposite among the two spacetime
components. Based upon the result at $-1/3$ we prove that the
ratio of brane tension to be one as expected. Furthermore, we
discuss the factorization of star algebra in the presence of B
field under zero-slope limit and comment on the implications of
our results to the recent proposed map of Witten's star to Moyal's
star. \noindent
\end{abstract}
\end{titlepage}

\setcounter{footnote}{0}

\section{Introduction}
  The non-perturbative physics of Witten's cubic string field
theory \cite{Witten,GJ1,GJ2} has been enhanced a lot since the
proposal of vacuum string field theory(VSFT) by Rastelli, Sen and
Zwiebach in \cite{RSZ1} to \cite{RSZ7}, see also the related works
\cite{GT1} to \cite{Okuyama2}. The basic idea is to assume the
universality of the BRST operator around the closed string vacuum
after tachyon condensation such that the ghost and matter sector
decompose and the classical solution of the matter string field is
a projector, or the so called sliver state and its tachyonic lump
generalization \cite{RSZ2}.

    These projector states are understood as the non-perturbative
D-brane soliton of the VSFT, so the ratio of the branes' tension
provide a nontrivial test of the VSFT proposal. The ratio is
mainly determined by the spectrum of the Neumann matrix with and
without the zero modes as following \cite{RSZ2,Okuyama2}
\be
\label{R1} R={T_p \over 2\pi \sqrt{\a^{'}} T_{p+1}}={3(V_{00}+{b\over
2})^2 \over \sqrt{2\pi b^3}}
{\det(1-M^{'})^{3\over4}(1+3M^{'})^{1\over 4}\over
\det(1-M)^{3\over4}(1+3M)^{1\over 4}}\;,
\ee
where $M$ is the Neumann
matrix in the oscillator basis without zero modes, and $M^{'}$ is
the one with zero modes. The ratio has been first tested
numerically to be almost one in \cite{RSZ1} and then proved to be
one analytically by Okuyama in \cite{Okuyama2}. Since the spectrum
of both $M$ and $M^{'}$ contain $-1/3$ so that the ratio blows up
and requires properly regularization related to the issue of twist
anomaly investigated in \cite{Hata1}.

The spectrum of the Neumann matrix $M$ is constructed in
\cite{RSZ8} which shows that it has an doubly degenerate
continuous spectrum except at $-1/3$ which is non-degenerate.
Later on, the explicit calculation of the spectrum of the Neumann
matrix $M^{'}$ is also carried out in \cite{Feng1}. The result
shows that there are additional discrete spectrum besides the
continuous one found in \cite{RSZ8}. However, it also leads to
further puzzle that $M^{'}$ has doubly degenerate eigenvectors at
$-1/3$ so that naively the ratio R could be zero instead of one as
expected. More detailed study of the density of states is required
to resolve the issue.

   On the other hand, Witten's cubic SFT is believed to be
background independent although there is no general proof for the
background independence except some general arguments based upon
BCFT in the context of VSFT \cite{RSZ3} to \cite{RSZ5}. It is then
an issue to check the background independence case by case. A
simple case is to turn on the constant background B field, some
discussions for Witten's SFT on this background have been
discussed in \cite{Kawano1,Kawano2,Sugino}, there they argue that
the string vertexs can be obtained by an unitary transformation
from the one without the B-field.  Although this is the case, it
is still interesting to understand the background independence
issue in VSFT where the non-perturbative D-brane physics can be
worked out.

The construction of the sliver state and the tachyonic lump state
in the constant B field background based upon
\cite{Kawano1,Kawano2,Sugino} is done in \cite{Bonora}, and the
ratio of the brane tension between $Dp$ and $D(p+2)$ branes is
given by
\be
\label{R2} {\cal R}={(DetG)^{1\over 4}\over (2\pi \sqrt{\a^{'}})^2}{T_p
\over T_{p+2}}=({\theta^2+12(V_{00}+{b\over 2})^2 \over
4\sqrt{2\pi b^3} (DetG)^{1\over8} })^2{\det(1-{\cal
M}^{'})^{3\over4}(1+3{\cal M}^{'})^{1\over 4}\over
\det(1-M)^{3\over2}(1+3M)^{1\over2}}\;,
\ee
where $G$ stands for the open string metric.
${\cal M}^{'}$ is the Neumann matrix with the zero modes in
the oscillator basis, and due to the B field there is mixing
between two coordinate directions so that the dimension of $\cMp$
is twice larger than the one of $M$. In order to determine the
ratio one needs to have the spectrum of $\cMp$, at least one
should make sure if $\cMp$ has a eigenvalue at $-1/3$ in order for
the ratio to be finite \footnote{The finiteness of the ratio is
more subtle since some regularization is required, moreover, in
order to precisely determine the density states of the continuous
spectrum.  However, $-1/3$ is at the end of the spectrum so that
we believe the match of the degeneracy of $-1/3$ for $M$ and
$\cMp$ is essential for the ratio to be well-defined.}.

  Another interesting issue to which our results will be relevant is the
generalization of the recent proposed map of Witten's star to
Moyal's star \cite{Douglas} to the case with zero modes in the
constant B-field. There the Moyal conjugate pairs are the Fourier
transform of the twist-even coordinate modes and twist-odd
momentum modes, the Fourier basis are the eigenvectors of $M$ or
$\cMp$ in our case, and the twist parity is defined with respect
to left and right part of the open string in Witten's cubic string
field theory. So we need to construct the complete orthogonal
eigenvector basis of $\cMp$ in order to find out the proper Moyal
conjugate pairs such that the proposed map makes sense. If so,
then Witten's cubic string field theory can be reformulated in the
language of the noncommutative field theory, for which the soliton
is just a projector.

In section 2, we will summarize the basic of the VSFT by setting
up our conventions and notations, and warm up by calculating the
$1+\cMp$ part of the ratio $\cR$. Section 3 contains the main
results of our paper in which we will calculate both the
continuous and discrete spectrum of the Neumann matrix. In section
4 we will conclude our paper by discussing the implication of our
results to the ratio $\cR$ and to the map of Witten's star to
Moyal star. In the appendix we compute the ratio $\cR$ explicitly
and show that it is exactly one.

\bigskip
   {\bf Note added}: Upon finishing our paper we find that there appears two papers
 \cite{Feng2} and \cite{Bonora2} dealing with the same topic, and the first one
has substantial overlap with ours.

\section{Neumann matrix in the constant B field background}
\subsection{3-string vertex in the constant B-field background}
To construct the sliver state which is D25-brane in VSFT, we need
to use the 3-string vertex for the star product in the constant B
field background \cite{Kawano1,Kawano2,Sugino}, which is given by
\bea
\label{3stringp} |V_3> &=& \int dp^{(1)}dp^{(2)}dp^{(3)} \delta
(p^{(1)}+p^{(2)}+p^{(3)}) \exp \left( -\frac{1}{2} \sum_{r,s\leq
3} [\sum_{m,n\geq 1} V^{rs}_{mn} a^{(r)\dagger}_m
\cdot a^{(s)\dagger}_n \right. \nonumber \\
& & \left. +2 \sum_{m \geq 1} V^{rs}_{m0} a^{(r)\dagger}_m \cdot
p^{(s)}+ V^{rs}_{00} p^{(r)}\cdot p^{(s)}]-\frac{i}{2}\theta
\sum_{r<s}p^{(r)}
\times p^{(s)} \right) \nonumber \\
& & (|p^{(1)}>\otimes |p^{(2)}>\otimes |p^{(3)}>)
\eea
where
\be
|p>=\frac{1}{(\pi)^{1/4}}\exp(-\frac{1}{2}(p\cdot p) +
\sqrt{2}(a^\dagger_0\cdot p) -\frac{1}{2}(a^\dagger_0 \cdot
a^\dagger_0)) |0>
\ee
satisfy $\hat{p}|k> =k |k>$. The dot products are with respect to
the closed string metric along directions without $B$ field and to
the open string metric $G_{\a\b}$ along directions with $B$ field.
The $\theta$ is the noncommutative parameter and the cross product
means the contraction with the anti-symmetric symbol
$\epsilon^{\a\b}$.

As shown in \cite{Kawano1,Kawano2,Sugino} the B field will change
the Neumann boundary conditions of the free open string to the
mixed one and introduce the Moyal-like phase in the string vertex
besides replacing the closed string metric by the open string one in
the B-field directions. Obviously one can decompose the above
vertex into the transverse part $|V_3^{\bot}>$ and the parallel
part $|V_3^{\|}>$ with respect to the $B$ directions.

  To construct the lower dimensional brane to which the transverse
directions is no longer translational invariant, we need to use
the 3-string vertex in the oscillator basis as the star product,
which can be obtained by integrating out the momentum in
(\ref{3stringp}),
\be
\label{Vper} |V_3^{\|}>=\left( \frac{{\cal
N}_{\theta}\pi^{1/2}}{(1+V_{00})}\right)^{d^\prime/2} \exp\left(
-\frac{1}{2}\sum_{r,s\leq 3} \sum_{m,n\geq 0} {\cal
V}^{rs}_{\alpha\beta, mn} a^{(r)\alpha\dagger}_m
a^{(s)\beta\dagger}_n \right) (|\Omega>\otimes |\Omega> \otimes
|\Omega>)\;,
\ee
where $V_{00}\equiv V^{rr}_{00}\;$.

  The relations between $\cV$ and $V$ is obtained in \cite{Bonora}
and we summarize the results in the following
\bea
\label{V1} {\cal V}^{\alpha\beta, rs}_{mn}&=&
G^{\alpha\beta}V^{rs}_{mn}- \frac{{\cal N}_{\theta}}{b}
\sum_{t,v=1}^3 V^{rv}_{m0}U^{\a\b,vt}V^{ts}_{0n}
\\
\label{V2} {\cal V}^{\alpha\beta, rs}_{0n}&=& \frac{{\cal
N}_{\theta} }{\sqrt{b}}\sum_{t=1}^3 U^{\a\b,rt}V^{ts}_{0n}
\\
\label{V3} {\cal V}^{\alpha\beta,
rs}_{00}&=&G^{\alpha\beta}\delta^{rs}-{\cal N}_{\theta}
U^{\a\b,rs}
\eea
where
\be
U^{\a\b,rs}\equiv G^{\alpha\beta}\phi^{rs}-i a
\epsilon^{\alpha\beta}\chi^{rs}\;,
\ee
with
\be
\cNt\equiv {8b(V_{00}+{b\over2})\over \theta^2det(G) +12
(V_{00}+{\b\over2})^2}\;, \qquad a \equiv {\theta \over
4(V_{00}+{b\over 2})}\;,
\ee
and
\be
\chi^{rs}=\left( \ba{ccc}
0&1&-1 \\
-1&0&1\\
1&-1&0 \ea \right), \hspace{5ex} \phi^{rs}=\left( \ba{ccc}
1&-1/2&-1/2 \\
-1/2&1&-1/2\\
-1/2&-1/2&1 \ea \right)
\ee

Instead of using $V^{rs}$ one defines the matrices $M^{rs}$ as
following
\be
M^{rs}=CV^{rs}\;,
\ee
where $C_{mn}=(-1)^n\delta_{m,n}, m,n \ge 1$ is the twist matrix.
The matrices $M^{rs}$ are commuting to each other so that it is
useful to solve the algebraic equation for the projector states.
Similarly, as shown in \cite{Bonora} the matrices
\be
{\cal M}^{'rs}=C^{'}\cV^{rs}
\ee
are also commuting to each other. The generalized twist matrix
$C^{'}_{mn}=(-1)^n\delta_{m,n}, m,n \ge 0$. Although there are
nine Neumann matrices, only three of them are independent, in the
following we will only focus on one of them, that is, $M\equiv
M^{11}$, and the one with zero modes in the constant B-field
background $\cMp\equiv {\cal M}^{'11}$.  These two are the only
matrices appear in the ratio of brane tension in (\ref{R2}), and
will be our focus to find out their spectrum.

\subsection{The form of $\cMp$}
In the following we write down the explicit form of $\cMp$ by
adopting the basis introduced firstly in \cite{RSZ2}. After some
calculations, we arrive
\bea
{\cal M}^{'\a\b}_{00}&=&G^{\a\b}(1-{\cal N}_{\theta})\;,
\\
{\cal M}^{'\a\b}_{0n}&=&-\sqrt{2\over b}{\cal N}_{\theta}(G^{\a\b}
<v_e|-i \epsilon^{\a\b}{2a \over \sqrt{3}} <v_o|)\;,
\\
{\cal M}^{'\a\b}_{mn}&=&G^{\a\b}(M_{mn}-{2\over b}{\cal
N}_{\theta}(|v_e><v_e|-|v_o><v_o|)) \nonumber
\\
&&+i\epsilon^{\a\b} {2a \over \sqrt{3}}{2\over b}{\cal N}_{\theta}
(|v_e><v_o|+|v_o><v_e|)\;, \qquad m,n\ge 1\;,
\eea
where the vectors $|v_e>$ and $|v_o>$ are the same as defined in
\cite{Feng1} and \cite{Okuyama2}.  Note that
$(|v_{e,o}>)^T=<v_{e,o}|$ and
\be
C|v_e>=|v_e> \;,\qquad C|v_o>=-|v_o>\;.
\ee
Explicitly,
\be
<v_e|=(0,{A_2\over \sqrt{2}},0,{A_4\over \sqrt{4}},0, \cdots)\;,
\qquad <v_o|=(A_1,0,{A_3\over\sqrt{3}},0,\cdots)\;,
\ee
where
\be
({1+i z\over 1-i z})^{1/3}=\sum_{n\in even} A_n z^n+i \sum_{n \in
odd} A_n z^n\;.
\ee

 Unlike the case without B field, the Neumann matrix $\cMp$
carries both space-time and level indices, and there are imaginary
part associated with nonzero $\theta$. Moreover, the mixing of the
space-time indices due to the nonzero $\theta$ will make the
diagonalization of the Neumann matrix non-trivial.

Notice that, the open string metric
\be
G_{\a\b}=(1+(2\pi\alpha^\prime B)^2)\eta_{\a\b}
\ee
is diagonal and proportional to $\eta_{\a\b}$. And in the expression of the
three-string vertex, the inner product is respect to this open string metric.
However, since
\be
[a^\a_m, a^{\b\dagger}_n]=G^{\a\b}\delta_{mn}
\ee
we can rescale either the oscillators or the metric such that we
have normalized metric and commutator. This will give us the
freedom to use the well-known results on sliver state construction
and spectrum analysis, and the only effect is the overall
normalization factor $(DetG)^{1\over2}$ appearing in the ratio
formula (\ref{R2}) and also in the normalization of the
corresponding 3-string vertex and sliver state. From now on, we
will work under the convention $G_{\a\b}=\eta_{\a\b}, [a^\a,
a^{\b\dagger}]=\eta^{\a\b}$ with $\eta^{\a\b}=\delta^{\a\b}$.

\subsection{The spectrum of $M$ and the ratio $\cR=1$ }
The spectrum of M has been calculated in \cite{RSZ8}, and the
result is that it has a continuous spectrum labelled by $k$ such that
\be
M|k>=M(k)|k>\;,
\ee
where
\be
M(k)=-{1\over 2\cosh{\pi k\over 2}+1}\;,
\ee
and the eigenvector
\be
|k>=(v_1^k,v_2^k,v_3^k,\dots)^T\;,
\ee
is generated by
\be
f_k(z)=\sum_{n=1}^{\infty}{v_n \over \sqrt{n}}z^n={1\over
k}(1-e^{-k \tan^{-1}z)}\;,.
\ee

The eigenvectors constitute a set of complete orthogonal basis and
\be
\int^{\infty}_{-\infty} {|k><k| \over \cK(k)}=1\;,
\ee
where
\be
\cK \equiv {2\over k} \sinh{\pi k\over 2}\;.
\ee
However, $k$ is not an twist definite eigenstate but
\be
C|k>=-|-k>\;.
\ee
Since $C$ commutes with $M$\footnote{However, C does not commute
with $M^{12}$ and $M^{21}$.} so that each eigenvalue has doubly
degenerate twist-definite states except for $|k=0>$ which is twist
odd \cite{RSZ8}.

Later on it is also useful to have the following facts
\cite{Hata2, Okuyama2} about $|v_{e,o}>$,
\be
\label{vk} <k|v_e>={1\over k}{\cosh{\pi k\over 2}-1 \over
2\cosh{\pi k\over 2}+1}\;, \qquad <k|v_o>={\sqrt{3} \over k}
{\sinh{\pi k\over 2} \over 2\cosh {\pi k\over 2}+1}\;,
\ee
\bea
<v_e|{1\over 1+3M}|v_e>={1\over4}V_{00}\;,
\\
<v_o|{1\over 1-M}|v_o>={3\over 4} V_{00}\;.
\eea

Since there is an eigenvalue of $M$ at $-1/3$ so that the ratio
$\cR$ is not well-defined unless $-1/3$ is also an eigenvalue of
$\cMp$ with multiplicity two. In order to calculate $\cR$ we need
to study the spectrum of $\cMp$, especially at $-1/3$, this 
will be done in the next section.

However, it is straightforward to calculate the ratio of the tension.
As for $Det(1-\cMp)$, based
upon the above expression of $\cMp$ and the facts
\bea
det \left( \ba{cc}
A &B \\
C& D \ea
\right) &=& detA det(D-CA^{-1}B)\;,
\\
det(1+|u><v|) &=&1+<v|u>\;,
\eea
one obtains
\bea
Det^{1/2}(1-{\cal M}^\prime) \over det(1-M) &=&
{\cal N}_{\theta}det\left( 1-{\cal N}_{\theta}
(1+\frac{4a^2}{3}){1 \over {1-M}}|\nu_o><\nu_o|\right) \nonumber \\
&=&
\frac{4b^2}{\theta^2+12(V_{00}+{b\over2})^2}\;.
\eea
when $\theta=0$, this recovers the result in \cite{Okuyama2}.

As for the calculation of $Det(1+3\cMp)$ part, one has to
introduce the regulator to overcome the difficulty brought by the
existence of $-1/3$ eigenvalue. It has been worked out in
\cite{Bonora2} and at last one find that the ratio of the brane
tension is exactly $1$, as expected. For completeness, the
detailed calculation is given in the appendix with a slightly
different way of calculating the $Det(1+3\cMp)$ from the one used
in \cite{Bonora2}.

\section{Finding the spectrum of $\cMp$}
In this section we will find out the spectrum of ${\cal M}^\prime$
by using the method in \cite{Feng1}.

\subsection{The eigen-equations}
Let's assume the form of the eigenstate as
\be
v=\left( \ba{c}
g_1\\
g_2\\
\int h_1(k)|k> \\
\int h_2(k)|k> \ea \right)
\ee
where $g_1, g_2$ are some complex numbers corresponding to the zero modes and
$h_1(k), h_2(k)$ are the coefficients of the expansion on the
$|k>$ basis. Note that due to the complex elements in $\cMp$, we
assume that $h_i(k)$ could be complex. Let $\lambda$ be the
corresponding eigenvalue, then $\cMp v =\lambda v$ encodes the
following relations:
\bea
\lam g_1&=&(1-\cNt)g_1-\sqrt{2\over b}\cNt (\int h_1(k)<v_e|k>
+\frac{2ia}{\sqrt{3}}
\int h_2(k)<v_o|k>)\;, \\
\lam g_2&=&(1-\cNt)g_2-\sqrt{2\over b}\cNt (\int h_2(k)<v_e|k>
-\frac{2ia}{\sqrt{3}}
\int h_1(k)<v_o|k>)\;, \\
\lam \int h_1(k)|k> &=& -\sqrt{2\over b}\cNt (g_1 |v_e>
+\frac{2ia}{\sqrt{3}}g_2|v_o>)
+ \int h_1(k)M(k)|k>   \nn \\
& & -\frac{2\cNt}{b}(|v_e> \int h_1(k)<v_e|k> -|v_o> \int h_1(k)<v_o|k>)\nn \\
& &+\frac{i4a\cNt}{\sqrt{3}b}((|v_e> \int h_2(k)<v_o|k> +
|v_o> \int h_2(k)<v_e|k>) \label{eigen1}\\
\lam \int h_2(k)|k> &=& -\sqrt{2\over b}\cNt (g_2|v_e>
-\frac{2ia}{\sqrt{3}}g_1|v_o>)
+ \int h_2(k)M(k)|k> \nn  \\
& & -\frac{2\cNt}{b}(|v_e> \int h_2(k)<v_e|k> -|v_o> \int h_2(k)<v_o|k>)\nn \\
& & +\frac{i4a\cNt}{\sqrt{3}b}((|v_e> \int h_1(k)<v_o|k> +|v_o>
\int h_1(k)<v_e|k>)\;.  \label{eigen2}
\eea

Define
\bea
\cCe[h_1(k)]=\int h_1(k)<v_e|k> \hspace{5ex} &\cCo[h_1(k)]=\int h_1(k)<v_o|k> \\
\cDe[h_2(k)]=\int  h_2(k)<v_e|k> \hspace{5ex} &\cDo[h_2(k)]=\int
h_2(k)<v_o|k> .
\eea
And from the first two relations, we can rewrite $g_1,g_2$ as
\bea
g_1&=&\frac{1}{1-\cNt-\lam}\sqrt{2\over b}\cNt (\cCe+\frac{2ia}{\sqrt{3}}\cDo)\\
g_2&=&\frac{1}{1-\cNt-\lam}\sqrt{2\over b}\cNt
(\cDe-\frac{2ia}{\sqrt{3}}\cCo)
\eea
Put the above expressions into the
Eqs.(\ref{eigen1},\ref{eigen2}), and using the expansions
\bea
|v_e>&=& \int \frac{<k|v_e>}{\cK}|k> \\
|v_o>&=& \int \frac{<k|v_o>}{\cK}|k>
\eea
one have
\bea
\lefteqn{\int (\lam -M(k))h_1(k) |k>} \nn \\
&=& \left\{c_1\cCe+ic_2\cDo \right\}\int \frac{<k|v_e>}{\cK}|k>
+\left\{c_3\cCo
+ic_2\cDe\right\}\int \frac{<k|v_o>}{\cK}|k> \\
\lefteqn{\int (\lam -M(k))h_2(k) |k>} \nn \\
 &=& \left\{ c_1\cDe-ic_2\cCo\right\}\int \frac{<k|v_e>}{\cK}|k>
 +\left\{-ic_2\cCe
+c_3\cDo\right\}\int \frac{<k|v_o>}{\cK}|k>
\eea
where
\bea
c_1&\equiv&-{2\cNt \over b}\cdot{1-\lam \over 1-\lam-\cNt} \nn \\
c_2&\equiv&-\frac{2\cNt}{b}\frac{2\cNt-1+\lam}{1-\cNt-\lam}\frac{2a}{\sqrt{3}} \nn \\
c_3&\equiv&(\frac{2\cNt}{b})(1-\frac{4a^2}{3}\cdot
\frac{\cNt}{1-\cNt-\lam}) \nn
\eea

 From the argument in \cite{Feng1}, one could expect
\bea
-\d(k-k_0)r_1(k)&=& -(\lam-M(k))h_1(k)+(c_1
\cCe+ic_2\cDo)\frac{<k|v_e>}{\cK}
+(c_3\cCo+ic_2\cDe)\frac{<k|v_o>}{\cK} \nn \\
-\d(k-k_0^\prime)r_2(k)&=& -(\lam-M(k))h_2(k)+(c_1
\cDe-ic_2\cCo)\frac{<k|v_e>}{\cK}
+(c_3\cDo-ic_2\cCe)\frac{<k|v_o>}{\cK} \nn
\eea
where $r_i(k)$ undetermined, with $r_1(k_0)=0, r_2(k_0^\prime)=0$.
Now solve $h_1(k), h_2(k)$ from the above two relations, and
applying the operations $\int dk <v_e|k>$ and $\int dk <v_o|k>$,
we obtain
\bea
\cCe&=&A_{ee}(c_1\cCe+ic_2\cDo)+B_{1e} \\
\cCo&=&A_{oo}(c_3\cCo+ic_2\cDe)+B_{1o} \\
\cDe&=&A_{ee}(c_1\cDe-ic_2\cCo)+B_{2e} \\
\cDo&=&A_{oo}(c_3\cDo-ic_2\cCe)+B_{2o}
\eea
where
\bea
A_{ee}(\lam)&\equiv&\int \frac{<k|v_e><v_e|k>}{{\cal K}\cdot (\lam - M(k))}=<v_e|\frac{1}{\lam-M(k)}|v_e>, \\
A_{oo}(\lam)&\equiv&\int \frac{<k|v_o><v_o|k>}{{\cal K}\cdot (\lam - M(k))}=<v_o|\frac{1}{\lam-M(k)}|v_o>, \\
B_{ie}&\equiv&\int \frac{\d(k-k_0)r_i(k)<v_e|k>}{\lam -M(k)} \label{Be}\\
B_{io}&\equiv&\int \frac{\d(k-k_0)r_i(k)<v_o|k>}{\lam -M(k)}\label{Bo}\\
\eea
It's easy to find that one can group the above equations into two
sets:
\bea
\label{set1}
(1-c_1A_{ee})\cCe-ic_2A_{ee}\cDo &=& B_{1e} \\
ic_2A_{oo}\cCe+ (1-c_3A_{oo})\cDo &=& B_{2o}
\eea
and
\bea
(1-c_3A_{oo})\cCo-ic_2A_{oo}\cDe &=& B_{1o} \\
ic_2A_{ee}\cCo+ (1-c_1A_{ee})\cDe &=& B_{2e} \label{set2}
\eea

\subsection{Continuous spectrum}
If the determinant factor
\be
Det \equiv (1-c_1 A_{ee})(1-c_3 A_{oo})-c^2_2A_{oo}A_{ee}
\ee
is not zero, then we can invert the above two sets of equation
(\ref{set1}) to (\ref{set2}) and obtain the formal solutions
\bea
\label{sol1}
\cCe&=&{B_{1e}(1-c_3A_{oo})+ic_2A_{ee}B_{2o} \over Det}\;, \\
\cCo&=&{B_{1o}(1-c_1A_{ee})+ic_2A_{oo}B_{2e} \over Det}\;,\\
\cDe&=&{B_{2e}(1-c_3A_{oo})-ic_2A_{ee}B_{1o} \over Det}\;,\\
\cDo&=&{B_{2o}(1-c_1A_{ee})-ic_2A_{oo}B_{1e} \over Det}\;.
\label{sol4}
\eea

In order to have nontrivial solutions, that is, the above $B$
functions are not all zero, we need to have $\lambda -M(k)=0$ at
$k=k_0=k^{'}_0\ne 0$ to cancel the zero of $r_{1,2}(k)$ at $k=k_0$.
In this case,  we will have a continuous spectrum
$\lambda=M(k_0)=-{1\over 2\cosh{\pi k_0\over 2}+1} \in [-1/3,0)$.
Explicitly,
\bea
\lam -
M(k)&=&M(k_0)-M(k)=-M^{(1)}(k_0)(k-k_0)-{1\over2}M^{(2)}(k_0)(k-k_0)^2+\dots\;,
\\
\nn &=& -{\pi\sinh{\pi k_0 \over 2} \over (1+2\cosh{\pi k_0\over
2})^2}(k-k_0)-{1\over2}{\pi^2+{\pi^2\over2}\cosh{\pi k_0\over
2}-\pi^2\sinh^2{\pi k_0 \over 2}\over (1+2\cosh{\pi k_0\over
2})^3}(k-k_0)^2+\dots
\eea
where $M^{(n)}(k_0)\equiv {d^n M \over dk^n}|_{k_0}$.  Note that
$M^{(1)}(k_0)$ is non-vanishing except at $k_0=0$, however,
$M^{(2)}(0)\ne 0$. Therefore, if $k_0= 0$ then $\lam -M(k) \sim
-{1\over 2}M^{(2)}(0)k^2$, otherwise $\lam -M(k) \sim
-M^{(1)}(k_0)(k-k_0)$, to the leading order.

   In this subsection we focus on the case of $k_0 \ne 0$ and
will discuss the special case $\lam=-1/3$($k_0=0$) in the next
subsection. Based upon the behavior of $\lam -M(k)$ near $k=k_0$
we will have three kind of choices for the behavior of $r_i(k)$
around $k=k_0$. The first one is to let $r_i(k)\sim d_i(k-k_0)$ for
$i=1,2$, then we have nonzero
\bea
\label{Be0} B_{ie}(k_0)&=&{-d_i(\cosh{\pi k_0\over 2}-1) \over k_0
(2\cosh{\pi \k_0 \over 2}+1)M^{(1)}(k_0)}\;,
\\
\label{Bo0} B_{io}(k_0)&=&{-d_i\sqrt{3}\sinh{\pi k_0\over 2}\over
k_0(2\cosh{\pi \k_0 \over 2}+1) M^{(1)}(k_0)}\;,
\eea
and the corresponding eigenvector is
\bea
v(k_0)=\left( \ba{c} \frac{\sqrt{2\over
b}\cNt}{1-\lam-\cNt}\frac{(1+{2(-3+4a^2)\cNt
\over3b}A_{oo})B_{1e}+i\frac{2a}{\sqrt{3}}(1+\frac{4\cNt}{b}A_{ee})
B_{2o}}{Det}\\
-\frac{\sqrt{2\over b}\cNt}{1-\lam-\cNt}\frac{(1+{2(-3+4a^2)\cNt
\over3b}A_{oo})B_{2e}-i\frac{2a}{\sqrt{3}}(1+\frac{4\cNt}{b}A_{ee})
B_{1o}}{Det} \\
\frac{(c_1-(c_1c_3-c_2^2)A_{oo})B_{1e}+ic_2B_{2o}}{Det}\frac{1}{\lam-M}|v_e>
+ \frac{(c_3-(c_1c_3-c_2^2)A_{ee})B_{1o}+ic_2B_{2e}}{Det}
\frac{1}{\lam-M}|v_o>-{d_1\over M^{(1)}(k_0)}|k_0> \\
\frac{(c_1-(c_1c_3-c_2^2)A_{oo})B_{2e}+ic_2B_{1o}}{Det}\frac{1}{\lam-M}|v_e>
+ \frac{(c_3-(c_1c_3-c_2^2)A_{ee})B_{2o}-ic_2B_{1e}}{Det}
\frac{1}{\lam-M}|v_o>-{d_2\over M^{(1)}(k_0)}|k_0>\ea \right).
\label{vge}
\eea
Note that $v(k_0)$ is degenerate with $v(-k_0)$ which is in
general different from $v(k_0)$ because
\be
B_{ie}(-k_0)=B_{ie}(k_0)\;, \qquad B_{io}(-k_0)=-B_{io}(k_0)\;
\ee
and $A_{ee}$ and $A_{oo}$ are also even functions of $k_0$.

The second choice is $r_1(k)\sim (k-k_0)$ but $r_2(k)\sim
(k-k_0)^2$ or higher power of $(k-k_0)$ so that $B_{1e}$ and
$B_{1o}$ are the same as the ones in (\ref{Be0}) and (\ref{Bo0})
but
\be
B_{2e}=B_{2o}=0\;.
\ee

In this case we can form a pair of "twist-definite" states from
$v(\pm k_0)$ as following,
\bea
v_{+-}(k_0)&=&{1\over 2}(v(k_0)+v(-k_0))|_{B_{2e}=B_{2o}=0} \nn\\
&=&{B_{1e}\over Det}\left( \ba{c} \frac{\sqrt{2\over
b}\cNt}{1-\lam-\cNt}(1+{2(-3+4a^2)\cNt
\over3b}A_{oo})\\
0 \\
(c_1-(c_1c_3-c_2^2)A_{oo})\frac{1}{\lam-M}|v_e>
\\
-ic_2\frac{1}{\lam-M}|v_o> \ea \right)- \left( \ba{c}
0 \\
0\\
{1\over 2M^{(1)}(k_0)}(|k_0>-|-k_0>) \\
0\ea \right)\;. \label{vpm}
\eea
and
\bea
v_{-+}(k_0)&=&{1\over 2}(v(k_0)-v(-k_0))|_{B_{2e}=B_{2o}=0} \nn\\
&=&{B_{1o}\over Det}\left( \ba{c} 0\\
i\frac{2a}{\sqrt{3}}\frac{\sqrt{2\over b}\cNt}{1-\lam-\cNt}
(1+\frac{4\cNt}{b}A_{ee})
\\
(c_3-(c_1c_3-c_2^2)A_{ee}) \frac{1}{\lam-M}|v_o> \\
ic_2\frac{1}{\lam-M}|v_e> \ea \right)- \left( \ba{c}
0 \\
0\\
{1\over 2M^{(1)}(k_0)}(|k_0>+|-k_0>) \\
0\ea \right)\;. \label{vmp}
\eea
Here the subscript indices $+$ or $-$ indicate the twist parity of the
third and fourth components with
respect to the twist matrix $C_{mn}$.
Note that both $v_{+-}$ and $v_{-+}$ are the "twist-definite"
eigenstates although their third and fourth components have
different twist parity, this implies that two spacetime directions
cannot have the same twist parity for each "twist-definite"
eigenstate.

The third choice is $r_2(k)\sim (k-k_0)$ but $r_1(k)\sim
(k-k_0)^2$ or higher power of $(k-k_0)$ so that $B_{2e}$ and
$B_{2o}$ are the same as the ones in (\ref{Be0}) and (\ref{Bo0})
but
\be
B_{1e}=B_{1o}=0\;.
\ee
We will then get the corresponding eigenvectors similar to
(\ref{vpm}) and (\ref{vmp}),
\bea
u_{-+}(k_0)&=&{1\over 2}(v(k_0)+v(-k_0))|_{B_{1e}=B_{1o}=0} \nn\\
&=&{B_{2e}\over Det}\left( \ba{c} 0\\
-\frac{\sqrt{2\over
b}\cNt}{1-\lam-\cNt}(1+{2(-3+4a^2)\cNt
\over 3b}A_{oo})
\\
ic_2\frac{1}{\lam-M}|v_o> \\
(c_1-(c_1c_3-c_2^2)A_{oo}) \frac{1}{\lam-M}|v_e>
 \ea \right)- \left( \ba{c}
0 \\
0\\
0\\
{1\over 2M^{(1)}(k_0)}(|k_0>-|-k_0>)
\ea \right)\;. \label{ump}
\eea
and
\bea
u_{+-}(k_0)&=&{1\over 2}(v(k_0)-v(-k_0))|_{B_{1e}=B_{1o}=0} \nn\\
&=&{B_{2o}\over Det}\left( \ba{c}
i\frac{2a}{\sqrt{3}}\frac{\sqrt{2\over b}\cNt}{1-\lam-\cNt}
(1+\frac{4\cNt}{b}A_{ee}) \\
 0\\
 ic_2\frac{1}{\lam-M}|v_e> \\
 (c_3-(c_1c_3-c_2^2)A_{ee}) \frac{1}{\lam-M}|v_o>
 \ea \right)- \left( \ba{c}
0 \\
0\\
0 \\
{1\over 2M^{(1)}(k_0)}(|k_0>+|-k_0>)
\ea \right)\;. \label{upm}
\eea
and it is easy to see that they are
independent set from (\ref{vpm}) and (\ref{vmp}); moreover, the
general state (\ref{vge}) is the linear combinations of the four
independent "twist-definite" eigenstates. We then conclude that
there are four "twist-definite" eigenstates for each eigenvalue
except $-1/3$.  This is a natural generalization of the results
for $M$ \cite{RSZ8} and $M^{'}$ \cite{Feng1} since our $\cMp$
involves two spatial directions, also we have four independent
parameters $B_{ie}$ and $B_{io}$ in the general eigenstate
(\ref{vge}).

If we define the inner product of the vectors $u, v$ as $u^\dagger v$,
then it is not hard to see that among the four independent eigenvectors,
the ones with different twist parities are orthogonal to each other.

\subsection{The $\lam =-1/3$ case}

Now, let us take $\lam=-1/3$ as an eigenvalue in the continuous
spectrum. Then we have $k_0=0$ such that $M(k_0)=-1/3$. From the
requirement that $r_i(k)/(k-M(k_0))$ should have no poles at
$k_0$, one can determine the form of the functions $r_i(k)$ up to
a scale factor. Taking the freedom of choosing the scale factor and
to the leading order, we set
\be
r_1(k)=r_1 k^2, \hspace{5ex} r_2(k)=r_2 k^2
\ee
where $r_{1,2}$ are arbitrary constants\footnote{There is a freedom to
choose the value of $r_{1,2}$. In this paper, if we don't let the $r_i$
vanishing,  then we just choose
$r_i=1$. Choosing one of $r_i$ vanishing equivalent to choosing higher
order of $k$.}
. And
\bea
B_{2e}&=&B_{1e}=0 \nn \\
r^{-1}_2B_{2o}&=&r^{-1}_1B_{1o}=-\frac{6\sqrt{3}}{\pi} \nn
\eea
Then we obtain
\bea
\cCe&=&\frac{ic_2A_{ee}B_{2o}}{Det} \\
\cCo&=&\frac{(1-c_1A_{ee})B_{1o}}{Det} \\
\cDe&=&\frac{-ic_2A_{ee}B_{1o}}{Det} \\
\cDo&=&\frac{(1-c_1A_{ee})B_{2o}}{Det}.
\eea
Note that all of ${\cal C,D}s$ are dependent on $A_{ee}$ explicitly but
dependent on $A_{oo}$ only through $Det$.

It is straightforward to get the eigenvector for the $\lam=-1/3$
eigenvalue. The explicit form is
\be
v=\left( \ba{c}
\sqrt{2\over 3b}\frac{6\cNt a}{4-3\cNt}\frac{iB_{2o}}{Det}(1+\frac{4\cNt}{b}A_{ee}) \\
-\sqrt{2\over 3b}\frac{6\cNt a}{4-3\cNt}\frac{iB_{1o}}{Det}(1+\frac{4\cNt}{b}A_{ee}) \\
\frac{ic_2B_{2o}}{Det}\frac{1}{\lam-M}|v_e> +
\frac{(c_3-(c_1c_3-c_2^2)A_{ee})B_{1o}}{Det}
\frac{1}{\lam-M}|v_o>-\frac{36r_1}{\pi^2}|k=0> \\
\frac{-ic_2B_{1o}}{Det}\frac{1}{\lam-M}|v_e> +
\frac{(c_3-(c_1c_3-c_2^2)A_{ee})B_{2o}}{Det}
\frac{1}{\lam-M}|v_o>-\frac{36r_2}{\pi^2}|k=0> \ea \right).
\label{vect}
\ee

In \cite{Feng1}, it has been argued that the $-1/3$ eigenvalue is
doubly degenerate, with two independent eigenvectors. At first
looking, it seems that we have only one eigenvector. This is not
the case. In fact, due to our freedom to choose $r_i$, we can set
either $r_1(k)$ or $r_2(k)$ to have higher power expansion in
terms of $(k-k_0)$ such that one of $B_{io}, i=1,2$ vanishes while
the other finite. This will not spoil the whole story. On the
contrary, we will find that the choice will lead to two
independent eigenvectors and the above one is just the
superposition of them.

First let $B_{1o}=0$, which could be achieved by choosing
$r_1(k)\propto k^3$ or higher power of $k$. In this case, the
above eigenvector turns to be
\be
v_{+-,{-1\over3}}={B_{2o}\over Det}\left( \ba{c}
\sqrt{2\over 3b}\frac{6i\cNt a}{4-3\cNt}(1+\frac{4\cNt}{b}A_{ee}) \\
0\\
ic_2\frac{1}{\lam-M}|v_e>  \\
(c_3-(c_1c_3-c_2^2)A_{ee}) \frac{1}{\lam-M}|v_o> \ea
\right)-\frac{36}{\pi^2}\left( \ba{c}
0 \\
0\\
0\\
|k=0>\ea \right)\;. \label{vect1}
\ee

On the other hand, one could choose $B_{2o}=0$ by chooseing
$r_1(k)\propto k^3$ or higher power of $k$. Then the eigenvector
has the form
\be
v_{-+,{-1\over3}}={B_{1o} \over Det}\left( \ba{c}
0\\
-\sqrt{2\over 3b}\frac{6i\cNt a}{4-3\cNt}(1+\frac{4\cNt}{b}A_{ee}) \\
(c_3-(c_1c_3-c_2^2)A_{ee})\frac{1}{\lam-M}|v_o>  \\
-ic_2\frac{1}{\lam-M}|v_e> \ea \right)-\frac{36}{\pi^2}\left(
\ba{c}
0 \\
0\\
|k=0> \\
0\ea \right)\;. \label{vect2}
\ee

It is not hard to see that the general eigenvector (\ref{vect}) is
just the superposition of $v_{+-,{-1\over 3}}$ and $v_{-+,{-1\over
3}}$.  Moreover $v_{+-,{-1\over3}}$ and $v_{-+,{-1\over3}}$ are
kind of twist definite states as mentioned in the $k_0 \ne 0$
case.

One may wonder if (\ref{vect1}) (\ref{vect2}) are really the
eigenvectors of ${\cal M}^\prime$ with eigenvalue $-1/3$. This
could be checked directly with the explicit form of ${\cal
M}^\prime$. Unlike the matrix $M$ which has only one twist odd
state $|k=0>$ at $-1/3$, our $\cMp$ is doubly degenerate at $-1/3$
in the continuous spectrum.  However taking into account the space
dimensions involved in our discussion, we find that the degeneracy
at $-1/3$ in the case with $B$ field is only half of the one in the
case without $B$ field. Nevertheless, the degeneracy of $\cMp$ at $-1/3$
matches the
one of $M$.

 A subtlety is about the discontinuity of $A_{oo}$ at $-1/3$
discovered in \cite{Feng1}. As shown there, $A_{oo}(({-1\over
3})^+)={3\over 4}ln27$ but $A_{oo}(({-1\over 3})^+)=-\infty$.
Since in our (\ref{vect1}) and (\ref{vect2}) $A_{oo}$ only appears
in $Det$ which blows up if $A_{oo} \rightarrow -\infty$, in this case
the vectors in (\ref{vpm}) and (\ref{vmp})are just reduced to
$(0,0,|k=0>,0)^T$ or $(0,0,0,|k=0>)^T$ which obviously cannot
satisfy the eigen-equations. So the ambiguity is lifted and we
should choose $A_{oo}(({-1\over 3})^+)$ instead of
$A_{oo}(({-1\over 3})^-)$; therefore the degeneracy at $-1/3$ is
two.

\subsection{Discrete spectrum}

When the determinant factor $Det$ vanish, then one could have nontrivial solutions
only if $B_{ie}=B_{io}=0, i=1,2$. The solutions of $Det=0$ correspond to the
so-called discrete spectrum. Being the function of $\lam$, the equation has
an integral form, which make it very hard to solve analytically. However,
what we are really interested in is to see  if $\lam=-1/3$ is an eigenvalue or not.
The reasons are twofold: on the one hand, from the form of the ratio of tension,
the $-1/3$ eigenvalue of ${\cal M}^\prime$ is essential to cancel the zero from
det(1+3M). In order to have the finite ratio, one should expect the degeneracy
at $\lam=-1/3$ of ${\cal M}^\prime$ match the one of $M$. On the other hand,
from the experience of ${\cal M}$ without $B$ field, $-1/3$ exists as an eigenvalue
in the discrete spectrum and make the degeneracy higher\cite{Feng1}. Therefore,
it is quite important to investigate the possibility of $-1/3$ as an eigenvalue
in the discrete spectrum. To this aim,
taking $A_{ee}=-(3/4)V_{00}, \lam=-1/3$, we have
\be
Det=\frac{\theta^2[(\theta^2+12(V_{00}+b/2)^2)^2-384A_{oo}V_{00}b(V_{00}+b/2)]}
{ (\theta^2+12(V_{00}+b/2)^2)^2\cdot (\theta^2+12V_{00}(V_{00}+b/2))}
\ee
Obviously, $\theta=0$ is a solution. In fact, this recovers the well-known result that
$\lam=-1/3$ belongs to the discrete spectrum, no matter what $b$ is, in the case of $\theta=0$.
(see \cite{Feng1}). Certainly it is possible for specified values of $\theta$, $Det(\lam=-1/3)=0$.
In this case, one find
\be
\theta^2=\sqrt{384A_{oo}V_{00}b(V_{00}+b/2)}-12(V_{00}+b/2)^2.
\ee
Since $\theta$ is real, we have to require
\be
8bA_{oo}V_{00}-3(V_{00}+b/2)^2 \geq 0
\ee
The above relation set the range of $b$, in which it is possible to find a $\theta$ such
that $\lam=-1/3$ is a discrete eigenvalue. However, one could expect that for
generic value of $\theta$, $\lam=-1/3$ does
not belong to discrete spectrum, although it exists in the continuous
spectrum.

We will not discuss carefully the discrete spectrum here. In \cite{Feng2}, the
careful discussion on discrete spectrum and eigenvector has been worked out.

\subsection{The spectrum of $\cM^{\prime 12}$ and $\cM^{\prime 21}$}

Due to the fact that the matrices $\cM^{\prime rs}$ enjoy the same property as
the usual matrices $M^\prime$, it is straightforward to obtain the spectrum
of $\cM^{\prime 12}$ and $\cM^{\prime 21}$, once we know the spectrum of $\cMp$.
More precisely,
since the matrices $\cM^{\prime rs}$ are commuting with each other and satisfy the relations
\be
\cMp+\cM^{\prime 12}+\cM^{\prime 21}=(\cMp)^2+(\cM^{\prime 12})^2+(\cM^{\prime 21})^2=1, \hspace{3ex}
\cM^{\prime 12}\cM^{\prime 21}=\cMp(\cMp-1),
\ee
they share the same eigenvectors and their eigenvalues satisfy
\bea
\lam^{12}(k)-\lam^{21}(k)&=&\pm \sqrt{(1-\lam(k))(1+3\lam(k))} \nn \\
\lam^{12}(k)+\lam^{21}(k)&=&1-\lam(k). \nn
\eea
From the above two relations, it's easy to read out the spectrum of $\cM^{\prime 12}$
and $\cM^{\prime 21}$.

\section{Discussions and Conclusions}
  In this paper we have calculated the spectrum of the Neumann
matrix with zero modes in the constant B field background. We find
both the continuous and discrete spectrum. The existence of the
discrete spectrum will set a constraint on $b$ for a given
$\theta$; moreover, for the generic $\theta$ there is no $-1/3$ in
the discrete spectrum. We will take this as a good point to the
finite and nonzero ratio of tension $\cR$.

The continuous one is similar to the one of $M$ but with four
degenerate twist-definite eigenstates at each eigenvalue
except at $-1/3$ where it is only doubly degenerate.  The doubling
of the degeneracy at each eigenvalue compared to the one of $M$ is
expected since the B field mixes two spatial directions and
doubles the dimensions of the eigen-space; moreover, the double
degeneracy at $-1/3$ is required for the the ratio of the tension
$\cR$ to be well-defined. However, to determine precisely the
ratio from the spectrum requires the details of the density of
states which is beyond the scope of this paper. On the other hand
there is an analytic way to determine the ratio $\cR$ to be one a
la the regularization method of Okuyama \cite{Okuyama2} as shown
in \cite{Bonora2}.  The agreement adds more weights to VSFT and
its background independence.

Another issue on which our results are concerned is the new
proposed map of Witten star to Moyal star \cite{Douglas} where the
Moyal pair is the Fourier transform of a twist-even coordinate and
the twist odd momentum.  The Fourier basis are the twist-definite
eigenstates of $M$. In \cite{MT} and \cite{Douglas} it is shown
that at the zero slope limit, the star algebra factorizes into two
subalgebras which was first proposed in \cite{Witten2}: ${\cal A}
\longrightarrow {\cal A}_0 \otimes {\cal A}_1$. The ${\cal A}_0$
corresponds to the subalgebra of zero-momentum sector and the
${\cal A}_1$ corresponds to the $C^*$-algebra of spacetime
functions. The 3-string vertex factorizes accordingly
\be
|V_3> \rightarrow |V_3^{(0)}> \otimes |V_3^{(1)}>,
\ee
where $<x_1|\otimes <x_2|\otimes <x_3||V_3^{(1)}>$ can be
identified as the the kernel for the usual Moyal product with zero
noncommutativity \cite{Douglas}.

In the presence of $B$ field, the directions without $B$ field
could be treated along the above way, while the directions with
$B$ field are a little different. It is easy to see that the
factorization of the star algebra still works. However, one should
replace commutative $C^*$-algebra ${\cal A}_1$ with the
noncommutative one.  The zero slope limit is the equivalent to
(set $b=2$)
\bea
{\cal V}^{\alpha\beta,rs}_{mn} &\rightarrow
&G^{\alpha\beta}V^{rs}_{mn} +{\cal O}(\epsilon^2)
\\
{\cal V}^{\alpha\beta,rs}_{0n} &\rightarrow &0+ {\cal O}(\epsilon)
\\
{\cal V}^{\alpha\beta,rs}_{00} &\rightarrow &G^{\alpha\beta}
(\delta^{rs}-\frac{16}{12+\theta^2}\phi^{rs})+i\epsilon^{\a\b}{4\theta
\over 12+\theta^2}\chi^{rs}\;,
\eea
and we can obtain the ${\cal A}_1$ part of the longitudinal
3-string vertex $|V_{3}^{\|}>$, it is
\be
{\sqrt{\pi} \over 2}{1\over 1+{\theta^2 \over
12}}\exp\left(\frac{-1}{2}({-4+\theta^2 \over
12+\theta^2})\sum_{r}a_0^{(r)\dagger}\cdot a_0^{(r)\dagger}
-({8\over 12+\theta^2})\sum_{r<s}a_0^{(r)\dagger}\cdot
a_0^{(s)\dagger} -({4i\theta \over
12+\theta^2})\sum_{r<s}a_0^{(r)\dagger}\times a_0^{(s)\dagger}
\right)|0>
\ee
which is exactly the same\footnote{It is up to an overall constant
factor $3\pi/4$.} as the 3-string vertex for the canonical Moyal
product(see (2.26) constructed in \cite{Douglas}). This confirms
that the spacetime algebra is a noncommutative one, with Moyal
product replacing the usual pointwise product.

    Moreover, it deserves to mention that: in the recent paper
\cite{Bonora2} they show that the zero mode of the tachyonic lump
state in the constant B field background is nothing but the usual
noncommutative soliton as a projector discovered in \cite{GMS}.
This is complimentary to our above observation that the low energy
zero mode of the SFT in constant B field background is governed by
the noncommutative $C^*$-algebra.

  It is interesting to generalize the proposed map of Moyal star
to Witten star in \cite{Douglas} to the 3-string vertex with zero
modes and with B field.  This requires the eigenvectors of
$\cMp$ to form a complete orthogonal set in order to obtain the
proper Moyal conjugate pairs which are the twist-definite
eigenvector by the Fourier transform based on these eigenvectors.
Future work are required to carry out the program.

In this paper we did find the twist-definite eigenvectors of $\cMp$
, however, for each twist-definite eigenvector the twist parity is
opposite among the two spatial components. This implies that we
need to choose opposite twist parity assignments for the Moyal
conjugate pairs between the two spatial directions parallel to
the B field, that is, if we choose the twist-even coordinate modes
and the twist-odd momentum modes as the Moyal conjugate pairs for
one parallel direction to the B field, we should choose the
twist-odd coordinate modes and the twist-even momentum modes as
the conjugate pairs for another parallel direction.  This kind of
the flip of the twist parity for the Moyal pairs in two spatial
directions is unexpected, and it deserves more study to understand
its physical implication to the equivalence between Witten's
string field theory and the generalized noncommutative field
theory.

\section*{Acknowledgment}
BC would like to thank ITP, Chinese Academy of Science for the
hospitality during his visit. FLL would like to thank Pei-Ming Ho
and Sanefumi Moriyama for helpful discussions,  Hsien-Chung Kao
for his kind support, and  the courtesy of the CTP at NTU, the
NCTS, and the CosPA project, Taiwan, he was supported by the NSC
grant No. NSC89-2112-M-032-002.

\section{Appendix: The calculation of the ratio $\cR$}

In this appendix, we show how to get the ratio of tension ${\cal
R}=1$, even in the case with $B$ field.

The formula for the ratio $\cR$ has been given in (\ref{R2}), and
in section 2.3, we have found
\be
{Det^{1/2}(1-\cMp)\over det(1-M)} = {4b^2 \over \theta^2+
12(V_{00}+{b\over2})^2}. \label{-M}
\ee
As for the part involving $(1+3\cMp)$, we need to use regulator as
in \cite{Okuyama2} since there are double degenerate eigenvectors
at $-1/3$ as shown section 3.3. First we can rewrite
\be
\label{3M1}
Det(1+3\cMp)=(4-3\cNt)^2Det(\eta^{\a\b}(1+3M+Q) + i\epsilon^{\a\b}P)
\ee
where
\bea
Q& =& -\a |v_e><v_e| +\b |v_o><v_o| \nn \\
P&=& \g(|v_e><v_e|+|v_o><v_o|) \nn
\eea
with
\bea
\a &=& 48(V_{00}+{b\over 2}) \over \theta^2+12V_{00}(V_{00}+{b\over2}) \nn\\
\b &=& 48 V_{00} \over \theta^2+12V_{00}(V_{00}+{b\over2}) \nn\\
\g &=& 8\sqrt{3} \theta \over \theta^2+12V_{00}(V_{00}+{b\over2})
\nn
\eea
It is easy to see that
\be
\label{abc}
\a\b +\g^2={4\over V_{00}}\b\;,
\ee
which will be useful later.

Further simplifying (\ref{3M1}) we get
\bea
&& {Det(1+3\cMp)\over det(1+3M)^2} \\
&=& (4-3\cNt)^2det(1+{1 \over 1+3M}Q)det\left(1+{1 \over 1+3M}(Q-
P{1 \over 1+3M+Q})P\right)\;,
\label{Det}
\\
&=& (4-3\cNt)^2(1-\a H_{ee})(1+\b
H_{oo})(1-(\a+q\g^2)H_{ee})(1+(\b-p\g^2)H_{oo}) \label{3M2}\;,
\eea
where
\bea
p&=&<v_e|{1 \over 1+3M+Q}|v_e> = {H_{ee}\over 1-\a H_{ee}} \nn \\
q&=&<v_o|{1 \over 1+3M+Q}|v_o> = {H_{oo}\over 1+\b H_{oo}} \nn
\eea
and
\bea
H_{ee}&=&<v_e|{1\over 1+3M}|v_e> =\frac{1}{4}V_{00} \\
H_{oo}&=&<v_o|{1\over 1+3M}|v_o>.
\eea
Note that $H_{oo}$ is not well-defined as shown in \cite{Feng1}
and it requires proper regularization. Here we just keep it
formally and will regularize it later.

  We then combine the 2st(3rd) and the 4th(5th) factors in the
(\ref{3M2}), and use (\ref{abc}) we get
\be
{Det(1+3\cMp)\over det(1+3M)^2} =(4-3\cNt)^2 \left(1-\a
H_{ee}+\b(1-{4\over V_{00}}H_{ee})H_{oo}\right)^2.
\ee
The factor before $H_{oo}$ is formally zero, however, we know
$H_{oo}$ is divergent. Using the proper regularization as done in
\cite{Okuyama2}, one has the final result as following
\be
\left(1-{4\over V_{00}}H_{ee}\right)H_{oo}=\frac{\pi^2}{12V_{00}}
\ee

In the end we get
\be
{Det(1+3\cMp)\over det(1+3M)^2} = \left( \theta^2 + 4\pi^2 \over
\theta^2+ 12(V_{00}+{b\over2})^2 \right)^2 \label{3M}
\ee

Combing with (\ref{3M}) and using the fact that $DetG=(1+({\theta
\over 2\pi})^2)^2$ which appears in the ratio formula (\ref{R2}),
we obtain
\be
{\cal R} =1
\ee
as expected.

\end{document}